\documentclass[a4paper, 11pt]{article}
\usepackage{epsfig}
\usepackage{rotating}
\usepackage{latexsym}
\newcommand{\be}{\begin{equation}}
\newcommand{\ee}{\end{equation}}
\newcommand{\ba}{\begin{eqnarray}}
\newcommand{\ea}{\end{eqnarray}}
\newcommand{\bi}{\begin{itemize}}
\newcommand{\ei}{\end{itemize}}

\newcommand{\<}{\langle} 
\renewcommand{\>}{\rangle}  

\newcommand{\la}{\label}

\begin{document}
\begin{titlepage}

\begin{flushright}
DESY 04-230
\end{flushright}
\vspace*{0.7in}

\begin{center}
{\Large\bf The spectrum of SU($N$) gauge theories in finite volume \\}
\vspace*{1.0in}
{Harvey B. Meyer\footnote{On leave from: 
Rudolf Peierls Centre for Theoretical Physics, 
University of Oxford, Oxford OX1 3NP, U.K.}
\\
\vspace*{.2in}
Deutsches Elektronen-Synchrotron DESY\\
Platanenallee 6\\
D-15738 Zeuthen\\
\vspace*{.04in}
harvey.meyer@desy.de
}
\end{center}

\vspace*{0.6in}

\begin{center}
{\bf Abstract}
\end{center}
We compute the spatial-volume dependence of the 
spectrum of 4D SU$(3\leq N\leq 6)$ gauge theories 
by lattice Monte-Carlo techniques.  Setting the scale with 
the string tension, the spatial volume is $L^3$
with $0.78{\rm fm}\leq L\leq 2.3{\rm fm}$.
The Euclidean `time' direction is kept large enough 
to be considered infinite
and the boundary conditions are periodic in all four dimensions.
We study the mixing of torelon pairs with the scalar and tensor
glueballs, using a $2\times2$ Hamiltonian based on large-$N$ counting rules.
Looking to the other symmetry channels, finite-volume effects on the glueball 
spectrum are already surprisingly small in SU(3), and
they become rapidly smaller as $N$ is increased: several low-lying
SU(6) states have no finite-volume corrections at the 1-2$\%$  level,
at least down to $L=0.9$fm.  We discuss the relation of this work
with analytic calculations in small and intermediate volume, and with
Eguchi-Kawai reduction in the planar limit.

\end{titlepage}

\setcounter{page}{1}
\pagestyle{plain}
%
\section{Introduction}\label{section_intro}
It has long been realised that the volume dependence of SU($N$) 
gauge theories is an important source of information on their dynamics. 
In very small spatial volumes $L\Lambda\ll 1$, perturbation theory is 
applicable due to asymptotic freedom and dimensional reduction effectively
occurs with respect to all three space dimensions 
(we assume periodic boundary conditions).  This idea was used 
in~\cite{luscher219} to derive an effective Hamiltonian for the 
zero modes of the
 pure gauge theory, which allows one to compute the `glueball'
spectrum analytically. The technique was then gradually extended
by taking into account tunneling between perturbative vacua
to reach volumes corresponding roughly to 0.5fm~\cite{Koller:1987fq}
(for a review, see~\cite{baal}). After pioneering small-volume Monte-Carlo 
calculations by Berg and Billoire~\cite{early}, 
the analytic results were found to be in agreement with numerical data
for the SU(2) gauge group in the much-improved
 calculations~\cite{berg_billoire,Kripfganz:1988bh} of the late eighties; 
numerical results for the SU(3) gauge group remained more limited~\cite{vohwinkel}.
Also in the presence of light quarks, finite-volume techniques have proven useful
to gain insight into the theory. An effective small-volume action was derived 
in~\cite{Kripfganz:1988bh} for the SU(2) gauge theory including the one-loop 
contribution of light quarks, and later
extended to intermediate volumes~\cite{Michael:1989he}. The existence of analytic
results in small volume, even at finite lattice spacing~\cite{vanBaal:1989fr}, 
was exploited in~\cite{GarciaPerez:1996hi} to study the scaling behaviour of 
various lattice gauge actions. More recently,
the so-called $\epsilon$-regime, defined by $m_\pi L\ll 1$, 
has been used as a testing ground for chiral perturbation theory, 
through the connexion with random matrix theory~\cite{wittig}.

At intermediate volumes, $L\simeq1$fm, non-perturbative effects
are an obstacle for purely analytic calculations. However, physical 
quantities of the finite-volume theory can be related to observables of the 
corresponding infinite-volume theory. The finite-volume correction to 
 the mass of a stable state is proportional to  $e^{-\frac{\sqrt{3}}{2}\bar mL}$ 
(where $\bar m$
is the mass gap) and the proportionality constant is nothing but the forward 
scattering amplitude of two of the lightest particles in infinite 
volume~\cite{luscher_finivol, luscher_houches}. 
Further, unstable states will mix with the 
scattering states of their decay products in a finite volume, and the mass
splittings inform about the phase shifts in infinite volume, and in particular
on the width of the resonance~\cite{luscher_finivol2, luscher91}. 
Similarly matrix elements in 
finite volume can be related to the `weak' decay rate of `kaons' in infinite 
volume~\cite{lu_le}.

Another point of view on gauge theories in finite-volume is obtained
from t'Hooft's large-$N$ expansion, where $N$ is the number of colours~\cite{hooft}.
Indeed, large-$N$ QCD is thought to be a theory of free mesons and glueballs
(while baryons have masses of order $N$~\cite{witten_baryons}); this picture is 
supported by arguments involving the large-$N$ counting rules for Feynman diagrams
and the assumption of confinement at $N=\infty$. From the general relations 
between finite and infinite observables cited above, one immediately infers
that finite-volume effects should become weaker as $N$ increases, since widths and 
scattering amplitudes are suppressed by a power of $1/N$. 
Conversely, while the low-lying SU($N$) glueball spectrum in large volume
has been shown to have $1/N^2$ corrections~\cite{teper,plb},
the finite-volume effects on energy levels can be used to determine
non-perturbatively the $N$-dependence of certain three- and four-point functions.
This work can be seen as a first attempt in this direction.
By comparison,  measuring these $n$-point functions on the lattice
in a straightforward way would be extremely hard.

The argument on the size of finite-volume effects outlined above
 fits in with the ideas of \emph{reduction} initiated by
Eguchi and Kawai~\cite{egu_kaw}: the SU($\infty$) lattice gauge theory 
defined on a single lattice hypercube of size $a^4$ has equivalent
loop equations for its Wilson lines to those of the corresponding  theory on an infinite 
lattice, and corrections come from the connected correlation
 of traced Wilson loops, which is suppressed  by $1/N^2$.
In fact the proof given by the authors trivially extends to 
any $\hat L^4$ lattice, $L=\hat L a$. Thus if the vacua of the theory, defined 
(a) on the $\hat L^4$ lattice and (b) on the infinite lattice, `choose' the same class 
of solutions, finite-volume effects on Wilson loop expectation values vanish 
at $N=\infty$.

The equivalence of the loop equations 
holds provided that the centre symmetry $Z(N)$ (which is promoted
to a global $U(1)$ symmetry at $N=\infty$) does not break spontaneously.
Strictly speaking, at finite $N$ such a spontaneous symmetry breaking cannot occur 
in finite volume. But the energy barrier between the different vacua allowed by 
the symmetry grows with $N$ so that tunnelling between them becomes completely 
suppressed at $N=\infty$.  This point can be made more precise.
In the Euclidean path integral formulation one may reinterpret one of the `small' spatial
dimensions as the Euclidean `time' direction, which fixes the temperature, 
$T = 1/L > T_c$. In this picture, two of the three spatial dimensions have size $L$  and one 
is infinite (or at least $\geq L$). The thermal tunnelling from one vacuum to the other
was discussed in~\cite{weng_luc_tep} in a semi-classical way. 
In this process two different vacua must coexist `temporarily', and  
a domain wall must appear between the confined and one of the $N$ deconfined vacua.
Once the `temperature' $1/L$ is large, perturbation theory becomes applicable. 
The domain wall tension between two $Z(N)$ vacua is given by~\cite{pisar}
\be
\sigma_w(T=\frac{1}{L}) = \frac{b_k(N,L)}{L^3}, 
\qquad b_k=k(N-k)\frac{4\pi^2}{3\sqrt{3\lambda(1/L)}};\qquad \lambda = g^2N.
\ee
If a domain wall orthogonal to the large dimension 
 separates two $k$-vacua, its energy is $\sigma_w L^2$. At temperature $T=\frac{1}{L}$,
the tunnelling rate is thus proportional to the Boltzmann factor $e^{-b_k(N,L)}$. 
We note that due to asymptotic freedom, even at fixed $N$ the Boltzmann factor is suppressed 
as $L\rightarrow0$, albeit more slowly than a power law.
To obtain the tunnelling rate this expression ought to be multiplied  by further factors 
that take into account the entropy of domain walls; however they are not expected
to depend strongly on $N$, and therefore we conclude that the tunnelling is exponentially 
suppressed in $N$. At large $N$ the exponent is large, and this justifies the 
semi-classical method and language used in the argument. 
If we now return to the original interpretation of the system as a finite-volume
one at zero temperature, we see that from the point of view of Monte-Carlo simulations, 
the centre symmetry along one of the dimensions of size $L$ 
is practically broken permanently at finite, large $N$. 
This conclusion does not depend on the size of the dimension left unspecified, 
in particular it holds for $L^4$ and $L^3\times\infty$ lattices.

It is well-known that if one takes $\hat L =1$  the 
spontaneous symmetry breaking does occur (in the sense outlined above)
with periodic boundary conditions, a fact that 
 led to the `quenched'~\cite{quenched} and `twisted'~\cite{twist} Eguchi-Kawai models.
In the former case, the path integral is constrained in a way that maintains
a uniform distribution of eigenvalues of the link matrices on the unit circle.
In the latter case, the symmetry breaking is prevented by  
twisted boundary conditions~\cite{hooft_twist}.
In fact, the connexion between the twisted reduced model and the ordinary 
gauge field theory is then even more precise: for the  choice of symmetric twist,
the reduced theory at finite-$N$ has  loop equations equivalent
 (up to the $O(1/N^2)$ corrections to 
factorisation) to those of the corresponding theory on a \emph{periodic} 
$\hat L^4$ lattice, with $\hat L_N=\sqrt{N}$. 
In particular, at large but 
fixed $N$,  $\<{\rm Tr} U_\mu\>=0$ and 
$\<({\rm Tr} U_\mu)^{\hat L_N}\>=0$, but the latter (a Polyakov loop in the field theory) 
will have a very long autocorrelation time at sufficiently weak coupling.
Therefore, as one takes the continuum limit, $N$ has to be increased 
 so as to maintain the physical size of the 
box $L=a\hat L_N$ in which the gauge theory is defined in order 
to converge to the expected theory.

To compare the spectrum of the (partially) reduced and the original theory,
connected two-point functions must be considered.
A Hamiltonian version of the reduction 
(on an $L^3\times {\bf R}$ lattice) was discussed in~\cite{neub_levine}
and arguments for the equality of the spectra in the reduced model 
and in the infinite-volume theory were given.
In essence, one of the arguments is that the reduction property still holds if a source
term for closed Wilson loops is added to the action; since this is a generating 
functional for glueball correlation functions, the finite-volume corrections 
to the spectrum must also vanish in the planar limit.
For the argument to be complete, however, the equality of the generating functionals 
must be shown to hold beyond the linear term in the source.

Factorisation implies  that the $N=\infty$ theory corresponds
to a classical theory. In the coherent state method~\cite{yaffe}, 
the spectrum could in principle be  obtained by expanding
the classical Hamiltonian (including the $O(1/N^2)$ corrections) 
to quadratic order around the point of phase space corresponding to the vacuum state. 

In this work we investigate by Monte-Carlo techniques 
the finite-spatial-volume spectrum of the SU(3), SU(4) and SU(6) gauge theories, for
$0.78{\rm fm} \leq L\leq 2{\rm fm}$. By keeping $L>1/T_c$, we will avoid that 
the system gets locked into a $Z(N)$ broken phase for long periods of Monte-Carlo time
(especially for $SU(4)$ and $SU(6)$).
We study separately the symmetry sectors where  `additional' states
occur and those where there are only `intrinsic' finite-volume effects on
the states already present in the infinite-volume theory. In particular
we examine the $N$ dependence of the finite-volume effects in both cases.

In the next section we describe in more detail 
the theoretical expectations on the behaviour of the finite-volume
spectrum of pure gauge theories. We present the lattice results for 
the SU(3) gauge group (section~\ref{section_su3}), and then SU(4) and SU(6) 
(section~\ref{sec:sun}). 
We finish with a summary of the results and a general discussion.
%
\section{Finite-volume and planar physics}\label{section_vol}
We consider a four-dimensional Euclidean quantum field theory defined on a 
spatial hypertorus with an infinite time direction, $V=L^3\times {\bf R}$.
In a quantum field theory admitting a mass gap $\bar m$, the leading finite-volume
effect on the spectrum normally corresponds to the emission, propagation `around the 
world'  and reabsorption of the lightest particle. A  formula corresponding 
to this picture was established 
to all orders in perturbation theory in~\cite{luscher_finivol}: the 
mass shifts behave as $e^{-\frac{\sqrt{3}}{2}\bar mL}$, and the formula is 
accurate when the exponent is large. A simple picture for the origin
of this mass shift is that the particle interacts via a Yukawa potential 
with its `replica' located by periodicity in $L\times L\times L$ cells.
It is now tempting to jump to the conclusion
that the spectrum only has exponentially small corrections. However the situation
changes qualitatively if a `new' light state appears
 at volumes where $L$ is still large 
compared to the inverse mass gap of the infinite-volume theory -- a non-perturbative 
effect. This is precisely the situation of pure SU($N$) gauge theories, where
such an extra state is present in a finite box but 
becomes a very massive and broad resonance which decouples in the infinite-volume limit.
We now review in some detail how this happens.

Due to the global $Z(N)$ symmetry in the SU($N$) gauge theory, 
the eigenstates of the transfer matrix can be
 classified according to their winding number. 
Let us encode the  $Z(N)$ transformation properties of an operator
or state by a triplet $(e_x,e_y,e_z)$~\cite{luscher_houches}. 
A glueball has $(0,0,0)$, a torelon in the 
$\hat x$ direction has $(1,0,0)$, the triplets combine additively for a direct
 product of states and the entries are defined modulo $N$; moreover, 
 charge conjugation connects the sector $k$ to $N-k$. In particular,
it turns out that the lightest state for $L<1.5$fm is the torelon, which
acquires the phase $e^{i2\pi/N}$ under a centre transformation. One can then imagine
a state made of two torelons winding the same cycle of the hypertorus in opposite
directions. Such an object is invariant under the $Z(N)$ symmetry, and therefore
becomes part of the glueball spectrum. The three resulting states (one for each 
spatial direction) can then be classified into the singlet $A_1^{++}$ and the doublet
$E^{++}$ of the cubic irreducible representations. In general there is a splitting 
between these two `extra glueball states'.

Apart from the simple torelon pairs, one could imagine that
  even more  states appear in the $(0,0,0)$ sector in small volumes. 
For SU($N$), $N$ torelons winding a cycle of the torus 
in the same direction transform trivially under the centre of the gauge group.
The real part of such operators couple to the $A_1{++}$ and $E^{++}$ sectors,
while the imaginary part belongs to the $T_1^{--}$ triplet. 
For $k\leq N/2$, two $(k,0,0)$ operators constitute another possibility to construct
glueball operators. One can also consider 
operators transforming non-trivially under two or three of the $Z(N)$ symmetries, 
$(1,1,0)$ and $(1,1,1)$; two of them, winding in opposite direction, again couple
to $(0,0,0)$ states.
However, in the absence of interactions these states are all significantly heavier
than the lightest two states in the representations we shall consider. In the following
we shall see no numerical evidence that they play a role in the dynamics of 
the low-lying spectrum.

At $L\simeq0.8$fm, the single-torelon mass
is significantly lighter than half the lightest 
scalar glueball mass in large volume. Therefore, in the absence of interactions, 
the lightest state in the $A_1^{++}$ (and $E^{++}$) channels would be a two-torelon
scattering state.  
The interaction potential between the two zero-momentum states is 
in general suppressed by their dilution in the transverse directions, 
$\Delta V\propto (\ell_\perp/L)^2$ in our case, where $\ell_\perp$ is the average
width of the `string'; but this is hardly a suppression at the volumes we are 
interested in. Also, (working in the basis of states with 
definite relative momentum)
the higher relative momenta $\frac{2\pi n}{L}$ will contribute
little to the ground state for $L<1$fm. 

Since the torelon mass varies roughly proportionally to $L$ and the glueball 
mass is only weakly dependent on $L$, 
 we expect the dynamics  of torelons and their mixing with an ordinary glueball
state  to determine the mass of the two lightest $A_1^{++}$ (and $E^{++}$) states
 in a certain range of volumes. It turns out that it corresponds roughly to 
$0.8{\rm fm}\leq L\leq 1.2{\rm fm}$. In that range, there is a large energy 
gap to the third state  in these two representations.
That means that to a good (and controllable) approximation, we can study
the system with a $2\times2$ effective Hamiltonian. 

In order to apply the stationary perturbation theory of quantum mechanics,
we need to choose a starting orthonormal basis which diagonalises an
`unperturbed' Hamiltonian. We suppose here that the full Hamiltonian of the SU($N$) 
gauge theory can be expanded in inverse powers of $1/N$:
\be
H(L,N)=\sum_{k=0}^\infty \frac{H_k(L)}{N^k}.
\ee
The existence of the t'Hooft limit  implies that $H_o$ has the same eigenvalues as
the Hamiltonian of the SU($\infty$) theory\footnote{The $SU(\infty)$ theory may
have additional, decoupled sectors of free hadrons.};
the latter, according to conventional wisdom, describes non-interacting colour singlets.
$H_o(L)$ is  our `unperturbed' Hamiltonian. If
$m_T(L,N)$ is the torelon mass and $m_G(L,N)$ the glueball mass
in the relevant symmetry channel, then the restriction of $H_o(L)$
to the vector space span by these two states reads
\ba
H_o(L) = \left(\begin{array}{c@{\quad}c}2m_T(L,\infty) & 0 \\         
                               0   & m_G(L,\infty) \end{array} \right)~.  \la{eq:hamil_o}
\ea
We now consider the full hamiltonian $H(L,N)$ in the same basis:
\ba
H_{2\times2}(L,N) = \left(\begin{array}{c@{\quad}c}E_{2T}(L,N) & h(L,N) \\         
                               h(L,N)   &m_G(L,N) \end{array} \right)  \la{eq:hamil2x2}
\ea
with
\be
 E_{2T}(L,N) = 2m_T(L,\infty)+v_T(L,N),\qquad  m_G(L,N) = m_G(L,\infty)+v_G(L,N).
\la{eq:E_2T}
\ee
The  diagonal matrix elements $v_T$ and $v_G$ contain the terms describing
`intrinsic' finite-volume corrections of the unperturbed states. 
For instance, in an effective low-energy description, 
the force between flux-tubes is attractive as a result
of the exchange of the lightest particle, a scalar glueball;
such a force is always attractive. 
The glueball state also undergoes `intrinsic' corrections,
proportional in leading approximation to 
$e^{-c\bar m L}$~\cite{luscher219}.
However, equally important are the off-diagonal elements describing mixing. 
Pictorially one can imagine the following process:
two flux-tubes undergo `fusion',  after which they 
 form a closed, contractible flux loop. In the flux-tube model~\cite{isgur},
 a glueball precisely corresponds to such a loop: at this point the uncertainty
principle (and possibly curvature effects) prevent the loop from shrinking to a point.

Let us now discuss the $N$ dependence of the various contributions to the 
Hamiltonian~(\ref{eq:hamil2x2}).
The finite-$N$ energy-shift $v_G$ corresponds to the usual $1/N^2$ correction to 
the mass of a colour-singlet state, along with the virtual emission `around the world'
and reabsorption of a colour-singlet by a colour-singlet state, 
which is also a $1/N^2$ effect. Similarly $v_T(L)$ contains $O(1/N^2)$ corrections
to the mass of each torelon, and
the `elastic' interaction of torelons encoded in $v_T(L)$ proceeds
through the exchange of a  colour-singlet, and therefore is also 
$O(1/N^2)$. $h(L)$ on the other hand, corresponds to the decay amplitude
of a colour-singlet into two colour-singlets, which is $O(1/N)$. A simple way to
see this is that the decay rate of a glueball above the two-torelon threshold 
(if the transverse dimensions were large) would be proportional to $h^2$ 
in this formalism, according to `Fermi's golden rule'. 
On the other hand, the width of colour-singlets in the pure 
gauge theory is $O(1/N^2)$, which leads to the announced conclusion.

Thus we expect, at sufficiently large $N$, 
the functions $v_G$, $v_T$ and $h$ to take the functional forms 
\ba
v_G(L,N) &\sim& \frac{1}{N^2}~ \bar v_G(L)\la{eq:v_G}  \\
v_T(L,N) &\sim& \frac{1}{N^2}~\bar v_T(L) \la{eq:v_T}   \\  
h(L,N)   &\sim& \frac{1}{N}~ \bar h(L)~. \la{eq:h}
\ea

On the lattice, we of course compute what corresponds to 
the two eigenvalues of $H_{2\times2}(L,N)$, $m_o(L,N)$ and $m_1(L,N)$. 
Suppressing everywhere the $(L,N)$ dependence, one has \vspace{0.2cm}
\ba
 m_o+m_1 &=& E_{2T} + m_G \la{eq:aver}\\
\left(m_o-m_1\right)^2 &=& \left(E_{2T}-m_G\right)^2 + 4h^2~\la{eq:split}
\ea
The information extracted from the lattice
is in general not sufficient to determine all three unknown functions
$v_G,~v_T$ and $h$. However the large-$N$ counting rules can help us out.

Firstly, according to the latter, 
\be
m_T(L,N)=m_T(L,\infty)+O(1/N^2). \la{eq:m_T}
\ee
For the glueball mass, similarly
\be
m_G(L=\infty,N)=m_G(L=\infty,N=\infty)+O(1/N^2), \la{eq:m_G2}
\ee
since in large volume there is no other light state
the glueball can mix with, and 
\be 
m_G(L=\infty,N=\infty)=m_G(L,N=\infty)\la{eq:m_G3}
\ee
if the planar theory has no finite-volume corrections. Therefore, 
using Eqs.~\ref{eq:E_2T}, \ref{eq:v_G}, \ref{eq:v_T},
\ref{eq:m_T}, \ref{eq:m_G2} and \ref{eq:m_G3}, Eq.~\ref{eq:aver} 
becomes the prediction
\be
m_o(L,N)+m_1(L,N)=2m_T(L,N)+m_G(L=\infty,N)+O(1/N^2) \la{eq:test}
\ee
that can easily be tested on the lattice, 
since all quantities are at a given $N$.

Secondly, the half-splitting of the two lattice states
provides an upper bound for  $h(L,N)$:
\be
|h(L,N)| \leq \frac{1}{2}|m_o(L,N)-m_1(L,N)|. \la{eq:bound}
\ee
For $L$ such that 
\be
 1  -  \frac{E_{2T}(L,N)}{m_G(L,N)} \ll O(1/N),\la{eq:cond1}
\ee
Eq.~\ref{eq:split} tells us that $h(L,N)$
saturates the upper-bound, up to small corrections.
Since $E_{2T}(L,N)$ and $m_G(L,N)$ are not known \emph{a priori}, it is more 
convenient to express the condition above by trading them for
the known quantities $2m_T(L,N)$  and $m_G(L=\infty,N)$:
\be
1-\frac{2m_T(L,N)}{m_G(\infty,N)}\ll O(1/N),\la{eq:cond}
\ee
(the left-hand side of~\ref{eq:cond1} and~\ref{eq:cond} only differ 
by $O(1/N^2)$ terms).
If $h(L,N)$  is found to be $O(1/N)$, the procedure used to extract $h$ 
is self-consistent. 
What is interesting, from the point of view of $1/N$ counting, is that in 
the volumes satisfying Eqn.~\ref{eq:cond}
the finite-volume mass shift is of order $1/N$.
Because in large volume glueballs are relatively heavy in units of the string tension,
it is easy to see that this condition will necessarily be satisfied 
at some intermediate box size. 
Thus there are circumstances where the
expectation that finite-volume mass shifts are suppressed by $1/N^2$ can be 
wrong. Analogous violations of the `$1/N^2$ correction rule' for the spectrum
of SU($N$) gauge theory will be discussed in section~\ref{sec:discussion}.

\paragraph{}In conclusion our expectations are the following. 
In most representations of the cubic group, where no degeneracies appear as 
the volume is varied, the energy levels have small deviations from their 
infinite-volume values. These corrections are of order $1/N^2$ and behave as 
$e^{-c\bar m(L)L}$
(as long as  the exponent is large). In those representations where
degeneracies appear, such as the $A_1^{++}$ and $E^{++}$ which contain torelon pairs,
an interesting  interplay between the parameters $L$ and $N$ appears. When $L$ is such 
that the relative separation of states is much larger than $O(1/N)$, 
the deviations of the glueball masses from their infinite-volume values are
 $O(1/N^2)$. But as the torelon mass comes within $O(1/N)$ of half the infinite-volume 
glueball mass, the finite-volume corrections to glueball masses are enhanced 
through mixing to $O(1/N)$. In parallel, as the torelon-pair energy falls below 
the lightest $A_1^{++}$ glueball, the mass gap $\bar m(L)$
is given by that energy and finite-volume shifts become more rapid, as a function of $L$.
%
%
\section{Lattice results for SU(3)}\label{section_su3}
Our lattice calculations employ the standard plaquette action
and the equally standard 1:4 combination of heatbath and over-relaxation 
sweeps is used for the update.
We calculate ground and excited state masses $m$ in the zero-flux sector 
$(0,0,0)$ from Euclidean correlation 
functions of zero-momentum operators 
using standard variational techniques~\cite{Luscher:1990ck}. 
We extract the string tension, $\sigma$, from the mass of a flux loop 
that closes around a spatial torus. We perform calculations
for a fixed value of the inverse bare coupling $\beta = 6/g^2 = 6.0$
corresponding to a lattice spacing $a \simeq 0.097$fm, if we use 
$\sqrt{\sigma}=440$MeV. The calculations are on
lattices ranging from $8^3 48$ to $24^3 24$, corresponding to
a spatial extent $0.78{\rm fm}\leq L \leq 2.3{\rm fm}$.
 Since some states become very light
on small volumes, the time direction is gradually 
increased from 24 to 48 lattice spacings as the spatial volume is reduced
in order to suppress `around the world' contributions 
(proportional to $e^{-m(L_t-t)}$) to the correlation function $C(t)$.
In the $\hat L=8$ case we checked that increasing $\hat L_t$ to 64 does
not alter the measured spectrum.

It is well-known that it is crucial to include sufficiently different 
operators in the variational basis that overlap significantly onto the 
low-lying spectrum, particularly when one is trying to extract more 
than one energy level.
From the discussion of the previous section, it should be clear that 
what is required in the $A_1^{++}$ and $E^{++}$ sectors are 
contractible Wilson loops \emph{as well as} pairs of spatial Polyakov loops winding 
around a cycle of the spatial torus in opposite directions. 
We measure them with zero relative momentum, since a crude estimate of the 
energy of two non-interacting torelons with one unit of momentum $2\pi/L$ 
each turns out to be well above the lightest $A_1^{++}$ state for the $L$ of 
interest\footnote{We reserve the name `adjoint Polyakov loop' for the case where
the Polyakov loops are created at the same position. The use of such operators was 
pioneered by Berg and Billoire~\cite{early}.}. We take the linear combinations
of the scattering states that belong to the $A_1^{++}$ and $E^{++}$ representations
of $O_h$. We note that the overlaps between the two-torelon and the Wilson loop
operators is substantial in the region $L\simeq1$fm (see below).
 In the other representations, our basis of operators consists only of 
appropriate linear combinations of contractible Wilson loops.
All our operators, the Wilson loops and the scattering state of Polyakov loops,
are constructed with several levels of smearing and blocking 
(see e.g.~\cite{fuzzy}). We note however that
one must be careful in a small volume not to include operators so non-local 
that they contain paths winding around a cycle of the torus.

In some of the simulations we use a 2-level algorithm~\cite{2leva} that 
reduces the variance of rapidly decaying correlators, thereby
allowing us to extract the mass at larger time-separation.
This is especially relevant in small volumes, 
where it turns out that our fuzzy operators do not necessarily have 
as good overlaps on the physical states as in large volumes. On the smallest volume
we notice a significant increase in the auto-correlation time
of the spatial Polyakov loop expectation value (see Fig.~\ref{fig:hist}; 
both of these runs were employing the ordinary algorithm).
This may be attributed to the proximity of $L$ to $1/T_c$. $aT_c\simeq 7.25$
according to~\cite{weng_luc_tep}, and it appears that 
tunnelling transitions are moderately frequent 
on the $8^3\times48$ lattice. In such a situation a possible criterion to
 split up the data into independent blocks is to require that the magnitude of the 
average Polyakov loop in each block be less than a prescribed value.
Our jackknife bins cover a Monte-Carlo time that is larger than the width of
 a typical peak on Fig.~\ref{fig:hist}.
The way we store the measurements~\cite{2leva} in the 2-level algorithm
allows us to study in detail
the dependence of the statistical errors on the jackknife bins;
we are confident that our statistical errors are not biased by more than $10\%$.

Table~\ref{tab:su3G} contains our data on the various cubic representations.
In the next section we study the lightest few states 
in the $A_1^{++}$ (`scalar') and $E^{++}$ (`tensor') channels
before we move on to the other representations.
\paragraph{Channels with the lightest states\\}
The evolution of the lightest $A_1^{++}$ and $E^{++}$ states is illustrated on 
Fig.~\ref{fig:a1_epp}. Although we are only really interested in the first two states,
 we monitor the level of the third excited state to estimate the mass gap to
the next states. We also plot the $T_2^{++}$ states which are degenerate with the
$E^{++}$ states in the large volume (and continuum) limit.

We observe that below 1.2fm, there are two states under 2GeV, rather than one; they are
the $A_1^{++}$ and $A_1^{++*}$ states. Meanwhile the $A_1^{++**}$ is approximately
 degenerate with the large-voume $A_1^{++*}$.

The straightforward interpretation, namely that the fundamental state becomes lighter, 
and that the $n^{\rm th}$ excited state becomes degenerate with the
large-volume $(n-1)^{\rm th}$ excited state ($n=1,2,\dots$),  would be hard
to understand in a simple way. A far more natural interpretation is the one
developed in the previous section, namely 
that one particular state, associated with the torelon pair, 
becomes light on the small volumes. It crosses
the level of the large-volume $A_1^{++*}$ state around $L=1.5$fm, 
and that of the large-volume $A_1^{++}$ level around $L=1$fm. Each level-crossing
is accompanied by a maximal mixing of the torelon pair with the glueball state. 

A fit to the lightest energy level
for $\hat L\geq 12$ was attempted, assuming the functional form
to $m(L)=m_\infty-\frac{C}{L}
\exp\left(-\frac{\sqrt{3}}{2}m_\infty L\right)$~\cite{luscher_finivol}. 
The fit is acceptable but 
we find that the factor $C$ is large ($670\pm150$) and driven 
by the point at $\hat L=12$, signalling that the glueball exchange `around-the-world' 
is probably not the dominant effect in that range of volumes
(such a large factor was already found in~\cite{gerrit}).
The two lightest states come closest together at $L=1$fm. 
In a two-state mixing model, the identity of the two states
at smaller $L$ is inverted: the ordinary glueball component of the first-excited
state is larger than that of the ground state, which is now more of the torelon-pair
nature. This crossing apparently happens when the curve $m_o(L)$ has an inflexion point.
This interpretation is supported by the fact that our two-torelon operators have 
significant overlaps\footnote{The overlap is defined as 
$\frac{\<O_1|O_2\>}{\sqrt{\<O_1|O_1\>\<O_2|O_2\>}}$.}
 with the usual scalar glueball operators (0.32 at $\hat L=11$, 0.67 at 
$\hat L=10$ and 0.55 at $\hat L=9$, choosing each time our best operator of each type).

A similar pattern is seen in the $E^{++}$ channel: the ground state is pushed down
by the presence of an extra light state for $L\leq1.2$fm. Below that $L$, the first
excited state is roughly degenerate with the fundamental $T_2^{++}$ state, which has
only small finite-volume corrections. The same mixing and crossing interpretation 
thus seems appropriate: on the smaller volumes it is the first-excited state which
has a glueball-dominated wave function.

\paragraph{Two-state mixing model\\}
At $\hat L=12$, we observe a state with an energy of almost exactly $2m_T(L)$, both
in the $A_1^{++}$ and $E^{++}$ sectors, and the other light state still has very
small finite-volume corrections with respect to the large-volume glueball masses.
However the situation changes quite rapidly for $\hat L\leq 11$.
Compared to other symmetry channels, the scalar and tensor glueballs masses
suddenly undergo  finite-size shifts larger by about a factor of 2.

The near-degeneracy of the two lightest states in the region 0.9---1.1fm
suggests the application of the  two-state mixing model. Fig.~\ref{fig:mbar}
shows the average mass of the fundamental and first excited states (separately in
the $A_1^{++}$ and $E^{++}$ sectors). The lattice data is compared to the energy level
$m_T(L)+m_G(\infty)/2$ that one expects if torelon interactions and intrinsic finite-volume
corrections to the glueball mass vanish (see Eq.~\ref{eq:test}).
The agreement is good for $L\geq0.9$fm, while 
at smaller $L$ the true average mass is pushed up compared to the naive estimate.
This provides evidence that the `intrinsic' effects are small in the region of interest,
though one could be
fooled if the mass shift for the torelon-pair has opposite sign to that of the 
finite-volume glueball mass, which would result in a large cancellation of these 
effects in the quantity considered.

Ignoring this possibility for the moment, if we now suppose 
(in the spirit of large-$N$ counting rules) that the `unperturbed' energy
levels are \emph{separately} well described by the values $2m_T(L)$ and $m_G(\infty)$,
we can estimate the mixing strength $h(L)$ through Eqn.~\ref{eq:split}.
 For $\hat L=10$ we find
$h(L)=h\simeq 0.15(2)m_G(\infty)$ (the quoted error is statistical). 
In the $E^{++}$ sector, the splitting is smaller than $|m_G(\infty)-2m_T(L)|$ 
in the 0.9---1.1fm region, and mixing can only split the states further apart.
Perhaps unsurprisingly we must conclude that $N=3$ is not yet large enough for the 
$N$-counting rules
of section 2 to be reliable in the extraction of the mixing strength $h$. Nevertheless 
the mass splitting, given in Table~\ref{tab:splitting}, provides an upper bound on $h(L)$.

\paragraph{Other representations\\}
In the other representations  the complications discussed above do not 
occur. We consider the $T_2^{++}$, $E^{-+}$, $T_2^{-+}$, 
$A^{-+}$ and $T_1^{+-}$ representations
(see Figs.~\ref{fig:a1_epp},~\ref{fig:other_g}). 
The $T_2^{++}$ state is the easiest case, because of its relatively small mass.
As mentioned above, its volume dependence is weak and smooth. The same holds for the 
$E^{-+}$, $T_2^{-+}$ and $T_1^{+-}$ states, whose masses increase monotonically
as  the volume becomes smaller. The increase is at the $10\%$ level at $L\simeq0.88$fm, 
as we discuss more quantitatively below.  Unfortunately
our calculation in the `pseudoscalar' channel $A_1^{-+}$ is not sufficiently good to 
allow us to make precise statements about its volume dependence, but it is clear that
it is the only one of these states that is becoming lighter -- its  mass seems to fall off
rather abruptly around 1fm. We remark that our operators had  poor overlap onto 
the state on the smaller volumes, whilst having satisfactory overlap on the large volumes.

%
\section{Larger $N$}\label{sec:sun}
We repeat the calculation of the spectrum for SU(4) and SU(6);
our method is entirely analogous to the SU(3) case. The lattice spacing is slightly
larger, $a=0.11$fm if we set the scale with the string tension, 
but prior knowledge~\cite{teper} indicates that the system is already in the 
scaling region at that lattice spacing. 
In the $A_1^{++}$ and $E^{++}$ sectors, we again include the two-torelon operators
along with the Wilson loops.
Zero-momentum scattering states  of two loops with ${\cal N}$-ality  $k=2$ 
were not included, although they could have a substantial overlap with one 
of the first excited scalar glueballs for $N\geq4$, 
if its wavefunction were well described
by a closed $k=2$ string. This possibility  has recently been suggested~\cite{plb},
but we leave it as a subject of investigation for the future.
The SU(4) and SU(6) spectra are given in Table~\ref{tab:su4G} and~\ref{tab:su6G}.
\paragraph{$N$-dependence of `intrinsic' finite-volume effects\\}
We start with the simpler case, namely the 
comparison of states not affected by the torelon pairs.
On Fig.~\ref{fig:other_rep} (top)  we compare the relative finite-volume mass shifts 
\be
\Delta(L)\equiv  \frac{m(L)}{m(L_\infty)}-1
\ee
of the lightest state between $L\simeq0.85$fm and $L=1.2$fm for the representations
$T_2^{++}$, $T_1^{+-}$, $E^{-+}$ and $T_2^{-+}$ ($L_\infty$ corresponds to our largest
lattice). Note that the error bars
on $m(L)$ and $m(L=\infty)$ are independent and therefore combine in quadrature.
We already observed in the previous section that the variations were small in SU(3), 
at the 10$\%$ level for $L\simeq0.88$fm.
Here we see that the finite-volume dependence gets even more suppressed 
 as one moves from SU(3) to SU(4) and SU(6). This suppression is rapid: for 
$N=6$ our data is consistent with vanishing finite-volume effects, in spite of 
relatively small error bars (1---2$\%$ for the $T_2^{++}$ and $T_1^{+-}$ cases).
Thus determining the  $N$ functional dependence numerically will require \emph{very}
  accurate data. On the other hand the finite-volume corrections vary quite smoothly
with $L$ in these representations, 
and therefore it is sufficient to compare different SU($N$) theories
at one chosen value of $\sqrt{\sigma} L\simeq 2$, for instance.
\paragraph{$N$ dependence of torelon-pair mixing with glueballs\\}
As we showed in the SU(3) case, the spectrum in the $A_1^{++}$ and $E^{++}$
sectors is affected by the presence of  torelon pairs. The average mass and the mass
splitting, that appear on the left-hand-side of Eqs.~\ref{eq:aver} and~\ref{eq:split} 
 are given in Table 4.  Recall that the half-splitting provides an upper-bound
for $h(L,N)$ (Eqn.~\ref{eq:bound}). 
Our data shows that the mixing strength is modest in magnitude but
the determination of its  $N$ dependence
requires accurate data for several 
$L$ in the region of maximal mixing given by Eqn.~\ref{eq:cond}.
Note that the minimum of the half-splitting as a function of $L$ gives $h$ 
only if the latter is $L$ independent.

Considering the bottom part of Fig.~\ref{fig:other_rep}, we see that the finite-volume
mass shifts, although smaller than for SU(3), 
 remain significant also for SU(4) and SU(6) in these representations, 
and that they are strongly
dependent on the precise value of $L$. Due to the crossing of states, the 
shift changes sign around 1fm (we used the state which is closest to $m_G(L=\infty)$
to compute the relative mass shift). These observations are consistent with the
expectations on the $N$ dependence of finite-volume effects formulated in section 2.
%
\section{Discussion}\label{sec:discussion}
We have studied the finite-size evolution of the low-lying glueball spectrum 
as a function of the number of colours in SU($N$) gauge theories. 
The box is $L^3$ in size, 
with $0.78{\rm fm}\leq L\leq 2.3{\rm fm}$, the boundary conditions are periodic
and the lattice spacing is fixed at roughly 0.1fm. Seven symmetry channels
were investigated.
In most of them ($T_2^{++},~E^{-+},~T_2^{-+},~T_1^{+-}$),
the states observed connect smoothly with their infinite-volume counterparts.
In the SU(3) case they become gradually more massive, and the mass shift is only at 
the $10\%$ level at 0.8fm. For SU(4) and SU(6) this mass shift is very suppressed,
and the SU(6) data is in fact consistent with vanishing finite-volume effects.
This constitutes the first observation of the large-$N$ suppression of these effects 
on the spectrum in Monte-Carlo simulations. To show that these effects are truly 
of order $1/N^2$ will require very accurate data. We also note that asymptotically
L\"uscher's formula~\cite{luscher_finivol, luscher_houches}
predicts a lowering of the glueball masses as $L$ decreases, 
since the exchange of the lightest scalar glueball yields
an attractive force; we conclude that the effect we are seeing is a different one, 
and that the asymptotic correction is too small to be observed with the present
statistical accuracy.

In the $A_1^{-+}$ `pseudoscalar' sector, the lightest glueball becomes significantly
lighter with respect to the infinite-volume state, and since this is observed for all 
gauge groups considered we think that this state
deserves a dedicated study. One might speculate that the mass shift
 is related to the lower average $Q^2$
in small volumes ($Q$ is the topological charge). In that case, one would also expect the 
pseudoscalar glueball to show up as a lighter state in the presence of light 
dynamical quarks, which suppress the topological charge fluctuations.
We note that a surprisingly sharp onset of the validity of random matrix theory 
in predicting the distribution of eigenvalues of the overlap Dirac operator in different
topological sectors on quenched configurations 
was observed around 1.1fm in~\cite{wittig}. It is not excluded that
the behaviour of the lightest pseudoscalar glueball is related to that observation.

The $A_1^{++}$ (scalar) and $E^{++}$ (tensor) sectors are most simply
interpreted in terms of  the appearance of extra light states, 
with respect to the large volume situation.
These have a natural interpretation in terms of pairs of flux loops winding
around one cycle of the hypertorus in opposite directions. Around 1fm the states
come close to the ground states in these sectors. A simple two-state mixing model
was used to interpret the two lowest energy levels.
It appears that the binding energy of the torelons and the intrinsic mass shifts
of the glueballs are rather small, so that the main cause of the variation of the spectrum
in the $L=1$fm region is mixing. In SU(3) it was estimated that the mixing energy amounts
to about $15\%$ of the infinite-volume glueball mass in the scalar sector.
This estimate still has a very large systematic uncertainty;
 the present study of this quantity must be viewed as exploratory.
However we note that in practice it is easy to tune $L$ so that Eqn.~\ref{eq:cond} holds;
one can thus 
expect to determine the $N$ dependence of the mixing energy $h$ in the near future.
Theoretical arguments were given in section 2 why the mixing energy should be of 
order $1/N$ and therefore the dominant effect at large $N$ in the cross-over region.

The issue of mixing 
of near-degenerate states in the large $N$ expansion is an important and general one.
For instance, a $k=2$ string~\cite{lucini} of length $L$ 
can be thought of as a state resulting from the 
mixing of two `unperturbed' states\footnote{They correspond to operators ${\rm Tr} \{P^2\}$ and 
$({\rm Tr} \{P\})^2$, if $P$ is the Polyakov loop.} 
which both  have $1/N^2$ corrections to their energy 
with respect to the planar limit value $2\sigma L$ ($\sigma$ is the fundamental string
tension). 
However a  mixing energy of order $1/N$ implies that the lightest $k=2$
string energy  deviates by a term of that order from $2\sigma L$~\cite{altes}.
Thus studying the mixing strength of torelon pairs with glueballs
in intermediate volumes satisfying Eqn.~\ref{eq:cond} addresses the same basic issue
concerning SU($N$) gauge theories in a  physical context where it is much easier for 
lattice calculations  to give a clear-cut answer.

To extend this work to  $L<1/T_c$, one has to confront the problem of long 
auto-correlation times on any quantity sensitive to the centre symmetry $Z(N)$.
This is especially true at large $N$, as outlined in the introduction. It then
appears to be more advisable to use twisted boundary conditions, as was done for SU(2) 
in the work of Stephenson and Teper~\cite{stephenson} following the 
analytic work~\cite{GonzalezArroyo:1988dz}. Working with larger gauge groups would 
then allow one to test the theoretical ideas of reduction. 
Eventually  volumes are reached 
where perturbation theory becomes applicable and accurate ($L\simeq 0.2$fm), 
so that the connexion with \emph{ab initio} analytic methods~\cite{luscher219}
can be established. It would thus be interesting to investigate
the spectrum of the effective small-volume Hamiltonian for the SU($N$) theory, $N$ large. 
A necessary and non-trivial condition to give agreement
with the large-volume spectrum, that is satisfied by this Hamiltonian even at finite $N$,
 is that it becomes spherically symmetric on very small volumes~\cite{luscher219}.
In view of the lessons learnt in this work and the fact that 
the multi-torelon operators decouple  from the glueball-creating operators at large $N$,
we would expect only certain energy levels  
to match those of the gauge theory in infinite volume.
%
\section*{Acknowledgements}
%
%
I am indebted to all participants of an informal seminar set up by 
M.~Teper and M.~Schvellinger
in Oxford, where several subjects touched upon in this article were discussed.
In particular I am grateful to H.~Vairinhos for his presentation of Eguchi-Kawai
reduction.
The numerical calculations were performed on a PPARC and EPSRC
funded  Beowulf cluster of the Rudolf Peierls Centre for Theoretical Physics.

\appendix
%
%
\begin{table}
\begin{center}
\begin{tabular}{c@{~~}c@{~~}c@{~~}c@{~~}c@{~~}c}
\hline 
\hline
IR& $\hat L=8$ & 9 & 10 & 11  & 12    \\
\hline 
torelon   &0.2107(14)&0.2599(30) & 0.3201(51) &0.4016(48)  & 0.4585(25)  \\
$A_1^{++ }$  & 0.54(1) &  0.53(1) & 0.55(1) &  0.63(1) & 0.661(9) \\
$A_1^{++*}$  & 0.731(8)& 0.77(2) & 0.77(2) & 0.88(3) & 0.93(2)  \\
$A_1^{++**}$    & 1.23(4) & 1.24(7) & 1.15(3) &{\small$<1.37(2)$}&1.24(5) \\
$E^{++}$        & 0.519(4)& 0.583(16) & 0.67(1) & 0.883(16)& 0.91(1)\\
$E^{++*}$       & 1.12(2) & 1.07(3)   & 1.04(1) & 0.97(3)  & 1.075(7)\\
$E^{++**}$      & 1.39(4) & 1.46(9)   & 1.42(4) & 1.54(12) & 1.46(7)\\
$T_2^{++}$      & 1.16(2) & 1.18(2)   & 1.12(1) & 1.127(5)& 1.06(1) \\
$T_2^{++*}$     & 1.46(6) & 1.59(8)   & 1.55(5) & {\small$<1.72(2)$} & 1.56(2) \\
$A_1^{-+}$&   1.07(3)& 1.04(3) & 1.19(3) &  1.20(3) & 1.25(5) \\
$E^{-+}$& {\small $<1.62(2)$} & 1.51(6) & 1.45(4)  & 1.42(5)& 1.39(6) \\
$T_2^{-+}$ & 1.52(7) & 1.53(5) & 1.35(3)& 1.44(4) & 1.41(1) \\
$T_1^{+-}$& 1.50(6) & 1.46(5) & 1.41(3) & {\small$<1.48(3)$} & 1.39(3) \\
\hline 
\hline
\end{tabular}\vspace{2.0cm}\\
\begin{tabular}{c@{~~}c@{~~}c@{~~}c@{~~}c}
\hline
\hline
IR &  $\hat L=$14  & 16 & 20  & 24  \\
\hline
torelon &  0.5739(34) & 0.6781(33) & 0.8860(64) & 1.082(60) \\
 $A_1^{++ }$ & 0.68(1)   &  0.697(4)  & 0.701(3)   & 0.698(2)  \\
$A_1^{++* }$ &  --  &  1.17(2) & {\small$<1.32(3)$} & {\small$<1.23(2)$} \\
$E^{++}$ & 1.06(2)&  1.060(6)& 1.07(1) & 1.073(3) \\
$E^{++*}$ & --  & 1.43(1)  &{\small$<1.56(4)$}  &{\small$<1.48(2)$}  \\
$T_2^{++}$ & -- & 1.070(5) &--  &  -- \\
$T_2^{++*}$ & -- &{\small$<1.50(1)$}& --  & --  \\
$A_1^{-+}$ & -- & 1.15(1) & -- &  -- \\
$E^{-+}$& -- & 1.36(1)  & -- & -- \\
$T_2^{-+}$ & -- & 1.37(1) & --  & --  \\
$T_1^{+-}$ & {\small$<1.42(2)$} & 1.31(2) & 1.31(2) & 1.34(2) \\
\hline
\hline
\end{tabular}
\caption{The SU(3) spectrum (in lattice units) in finite volume from Monte-Carlo simulations
at $\beta=6.0$ ($a\simeq0.097$fm, $V=L^3\times L_t$). $\hat L_t$ is 24 for $\hat L\geq 12$,
32 for $\hat L=9$ and 11, and 48 for $L= 8$ and 10. The method is described at the 
beginning of section~\ref{section_su3}. }
\label{tab:su3G}
\end{center}
\end{table}
%
%
\clearpage
\begin{table}
\begin{center}
\begin{tabular}{c@{~~}c@{~~}c@{~~}c}
\hline 
\hline
IR          &$\hat L=8$ & 10 & 16     \\
\hline 
torelon     &  0.2555(45)     & 0.4514(40)     & 0.8610(53)  \\
$A_1^{++ }$ & 0.4859(69)      & 0.6670(87)     &  0.729(10)  \\
$A_1^{++*}$ & 0.7297(97)      & 0.981(12)      &  1.338(13)  \\
$E^{++ }$   & 0.5465(71)      & 0.893(14)      &  1.139(11)  \\
$E^{++*}$   & 1.1342(78)      & 1.163(11)      &  1.606(16)  \\
$T_2^{++ }$ & 1.1682(68)      & 1.152(10)      &  1.143(16)  \\
$T_2^{++*}$ & 1.370(30)       & 1.599(22)      &  1.622(16)  \\
$A_1^{-+}$  & 0.924(14)       & 1.060(72)      &  0.991(86)  \\
$E^{-+ }$   & 1.528(18)       & 1.503(18)      &  1.507(50)  \\
$T_2^{-+}$  & 1.525(37)       & 1.477(40)      &  1.423(41)  \\
$T_1^{+-}$  & 1.491(20)       & 1.439(21)      &  1.453(15)  \\
\hline
\hline 
\end{tabular}
\caption{The spatial volume dependence of SU(4) 
glueball masses in lattice units at
 $\beta=10.90$ ($a\simeq0.11$fm, $V=L^3\times L_t$). $\hat L_t=24$ for 
$\hat L=16$, $\hat L_t =36$ for $\hat L=8$ and 10.}
\label{tab:su4G}
\end{center}
\end{table}
\begin{table}
\begin{center}
\begin{tabular}{c@{~~}c@{~~}c@{~~}c@{~~}c}
\hline 
\hline
IR          &$\hat L=8$ & 10 & 16 & 20    \\
\hline 
torelon     &  0.3096(38)  & 0.5016(63)  & 0.930(10)  & 1.2101(85) \\
$A_1^{++ }$ &  0.494(10)   & 0.707(11)   & 0.742(17) & 0.737(18)\\   
$A_1^{++*}$ &  0.782(11)   & 1.063(24)   &1.343(24) &1.315(33) \\
$E^{++ }$   & 0.589(12)    & 0.942(27)   &1.153(18) & 1.180(12)\\
$E^{++*}$   & 1.160(14)    & 1.175(17)   &-- & --\\
$T_2^{++ }$ & 1.1786(88)   & 1.177(12)   &1.167(13)  & 1.160(10)\\
$A_1^{-+}$  & 0.941(55)    & 1.203(26)   &$\leq$1.127(51)  & $\leq$1.293(19) \\
$E^{-+ }$   & 1.532(68)    & $\leq$1.573(28)   &1.498(21) & $\leq$1.61(2)\\
$T_2^{-+}$  & 1.532(18)    & 1.496(26)   &1.498(26) & 1.521(16)\\
$T_1^{+-}$  & 1.487(25)  & 1.462(27)  &1.474(17)  &1.47(2) \\
\hline
\hline 
\end{tabular}
\caption{The spatial volume dependence of SU(6) 
glueball masses in lattice units 
 at $\beta=25.05$ ($a\simeq0.11$fm, $V=L^3\times L_t$). $\hat L_t=24$ for 
$\hat L=16$ and 20, $\hat L_t =36$ for $\hat L=8$ and 10.}
\label{tab:su6G}
\end{center}
\end{table}
\begin{table}
\begin{center}
\begin{tabular}{c@{~~}c@{~~}c@{~~}c@{~~}c@{~~}c@{~~}c}
& IR          &$\hat L=8$ & 9 & 10 & 11 & 12     \\
\hline 
&  $A_1^{++}$: $m_+$ & 0.6394(44)  &0.648(18)  & 0.66(2)  & 0.760(17) & 0.792(13)\\
SU(3) & $\qquad m_-$  & 0.1708(80)  &0.248(18)  & 0.22(2) &0.255(30) &0.279(21)\\
\hline 
& $E^{++}$: $m_+$ & 0.793(21)  &0.794(37)  & 0.855(14)  & 0.918(35) &0.9935(53)   \\
&  $\qquad m_-$ & 0.6182(93)  & 0.478(28) &0.37(1)   & 0.120(20) & 0.177(10)  \\
\hline 
\hline 
& $A_1^{++}$: $m_+$   & 0.601(11)  &  & 0.8239(80)  &  &   \\
SU(4) & $\qquad m_-$            & 0.237(8)   &  & 0.314(13)  &   &  \\
\hline 
& $E^{++}$: $m_+$   & 0.8415(76)  &  & 1.0268(75) &  &   \\
&  $\qquad m_-$         & 0.575(15)  &  & 0.263(12)  &  &   \\
\hline 
\hline 
& $A_1^{++}$: $m_+$   & 0.6379(81)  &  & 0.885(14)  &  &   \\
SU(6) & $\qquad m_-$  &  0.285(12)   &   &0.355(24)   &   &  \\
\hline 
& $E^{++}$: $m_+$   & 0.8749(79)  &  &1.062(20)  &  &   \\
& $\qquad m_-$     & 0.582(28)  &  &0.241(36)  &  &   \\
\hline 
\hline 
\end{tabular}
\caption{The averages masses $m_+=(m_0+m_1)/2$ of and mass gaps $m_-=m_1-m_o $
 between the fundamental and first-excited states in the 
$A_1^{++}$ and $E^{++}$ sectors (in lattice units). Statistical correlations are taken into account
in the standard way by the jackknife method for $m_+$; for $m_-$, we give the mass 
plateau of $C_1(t)/C_o(t)$, where $C_o(t),~C_1(t)$ are the correlators of respectively 
the fundamental and first-excited  operators obtained from the 
variational method~\cite{var}. Its variance is again determined by the jackknife method.}
\label{tab:splitting}
\end{center}
\end{table}
\begin{figure}[t]
\centerline{\psfig{file=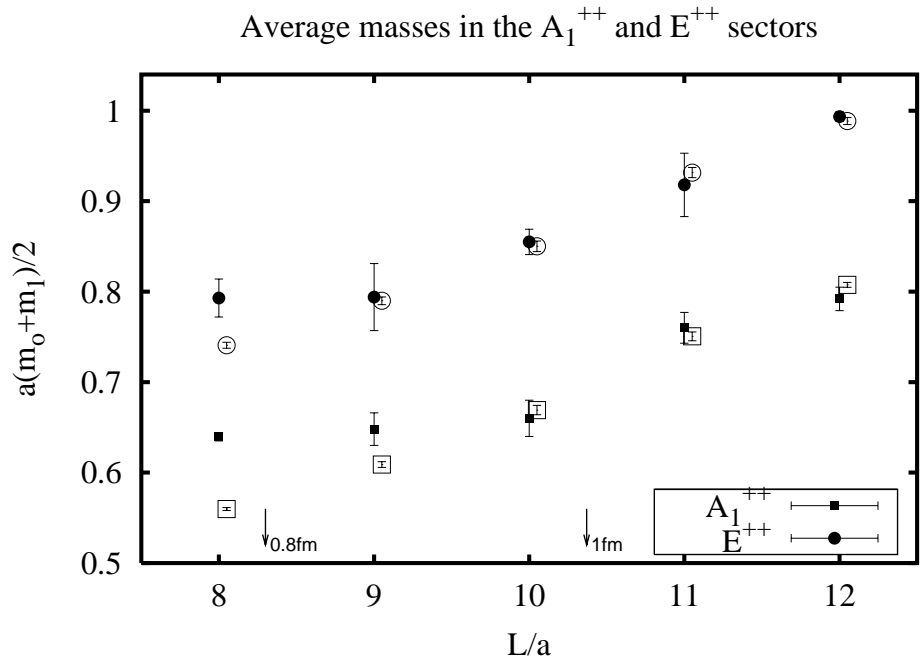,angle=0,width=13cm}}
\caption[a]{Average of the fundamental and first excited states in the $A_1^{++}$ and 
$E^{++}$ sectors (black symbols). 
This energy level is independent of two-state mixing. 
Also plotted (white symbols) is $m_T(L)+m_G(L=\infty)$. 
Where black and white symbol overlap, `intrinsic' finite-volume effects are small.} 
\la{fig:mbar}
\end{figure}
%
\clearpage
%
\begin{figure}[t]
\centerline{\begin{minipage}[c]{15cm}
   \psfig{file=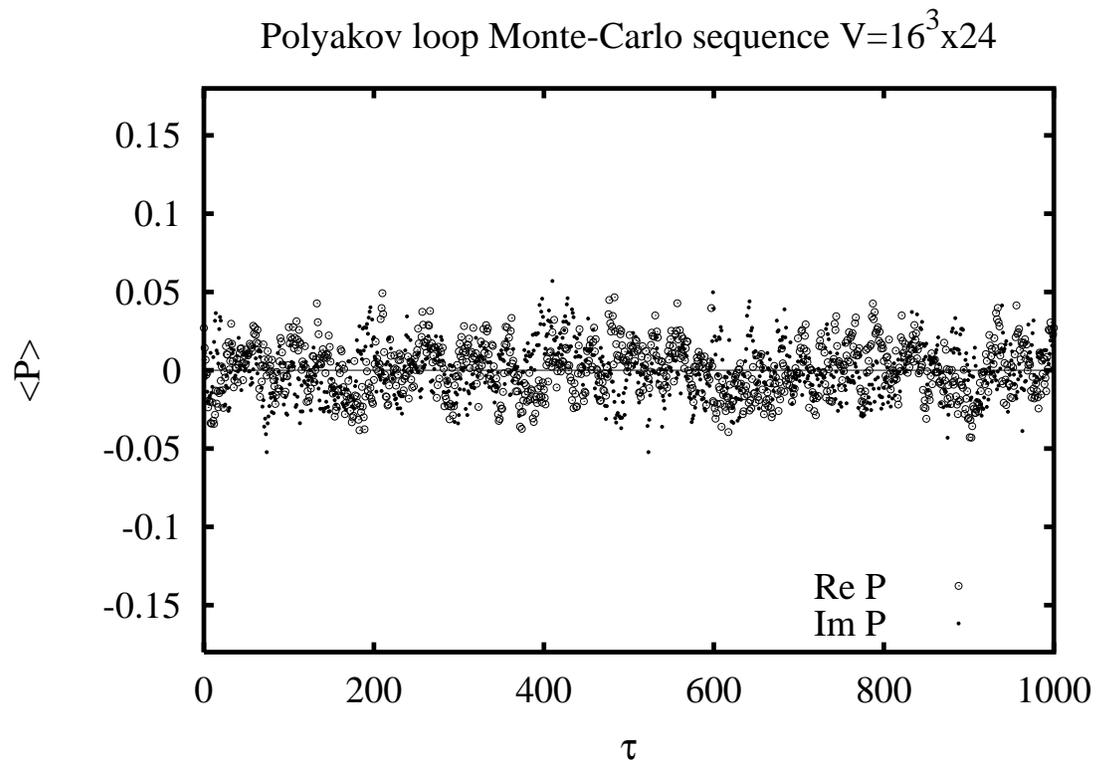,angle=0,width=15cm}
        \psfig{file=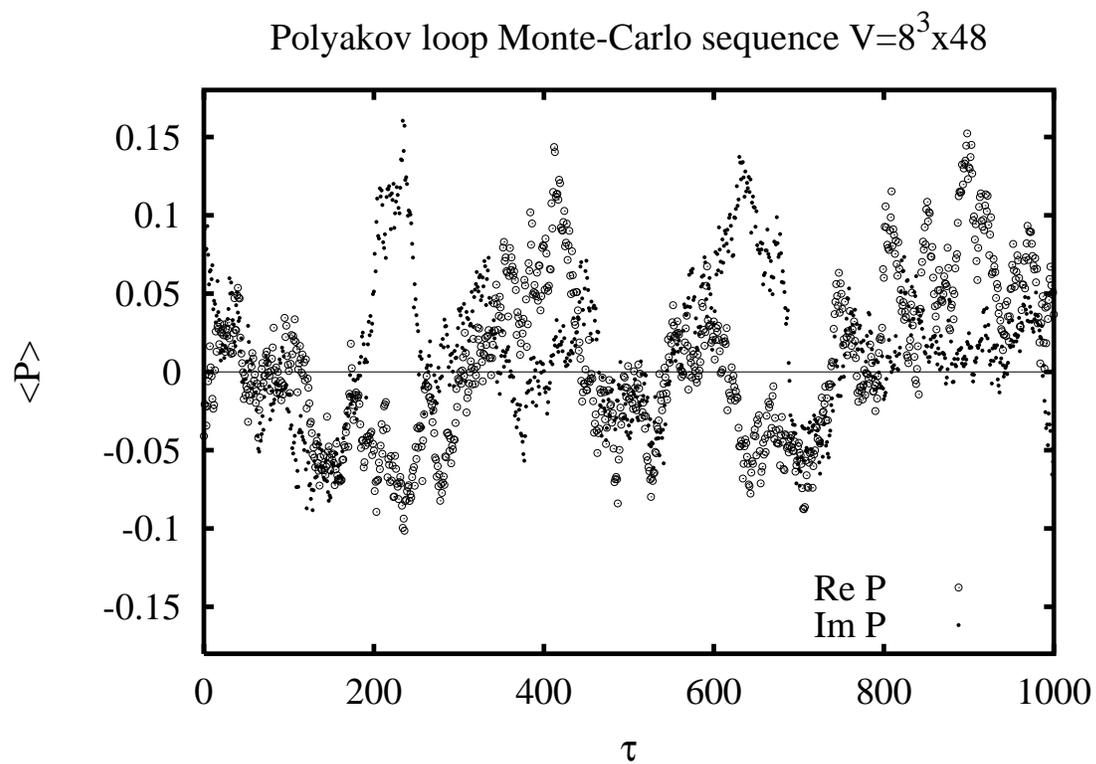,angle=0,width=15cm}
            \end{minipage}}
\caption[a]{Monte-Carlo sequence of the average value of the spatial-torelon operators
over $10^3$ measurements; the measurements are separated by 4 sweeps.}
\la{fig:hist}
\end{figure}
\clearpage
\begin{figure}[t]
\centerline{\begin{minipage}[c]{15cm}
   \psfig{file=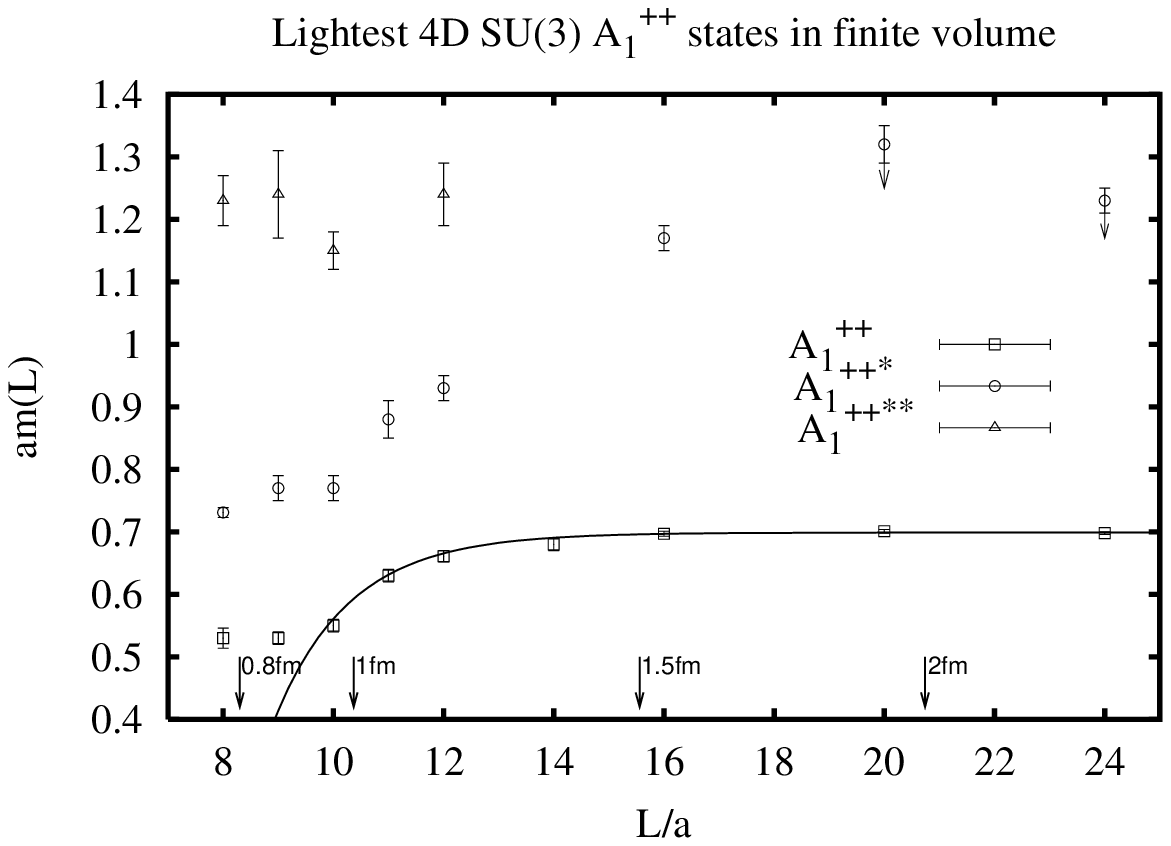,angle=0,width=15cm}
        \psfig{file=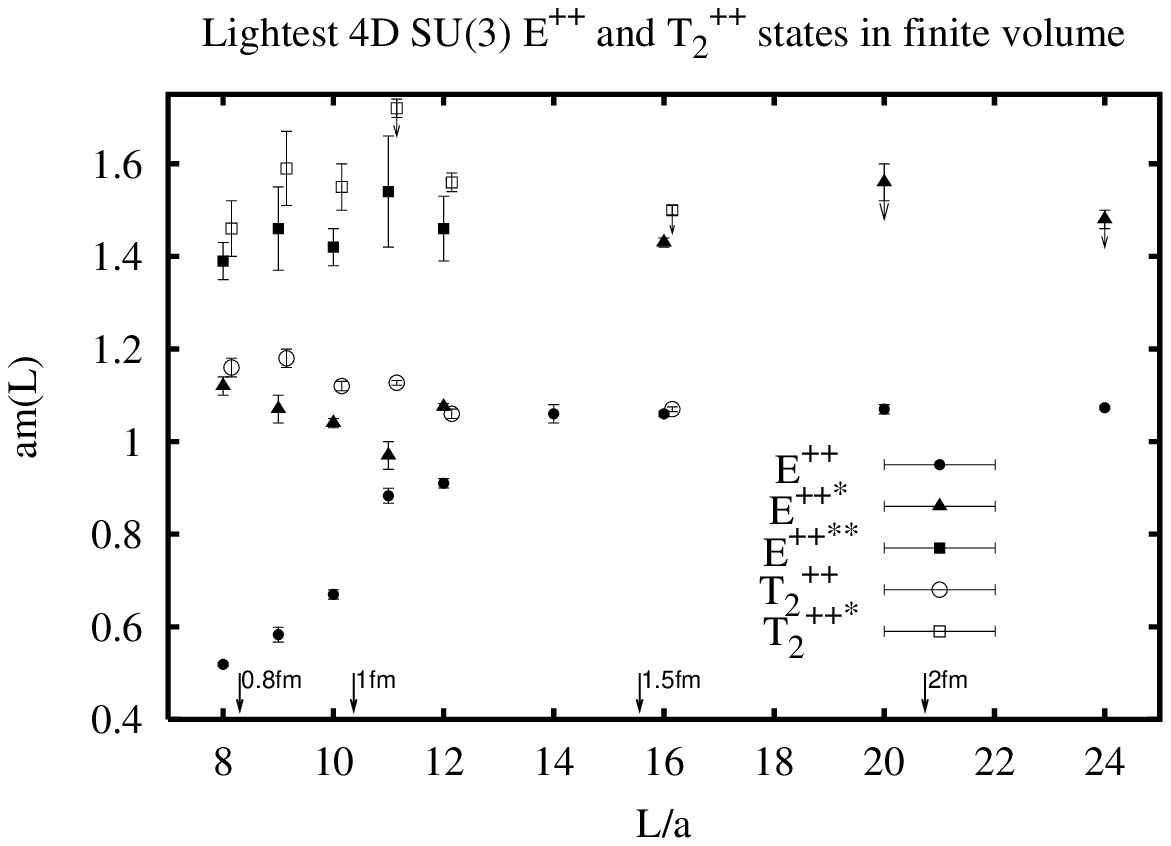,angle=0,width=15cm}
            \end{minipage}}
\caption[a]{Volume dependence of the lightest states in the 
representations that mix with torelon pairs. In the top plot, 
the smooth curve is a fit($0.699-674\exp(-0.605L)/L$) discussed in the 
text.}
\la{fig:a1_epp}
\end{figure}
\clearpage
\begin{figure}[t]
\centerline{\begin{minipage}[c]{15cm}
   \psfig{file=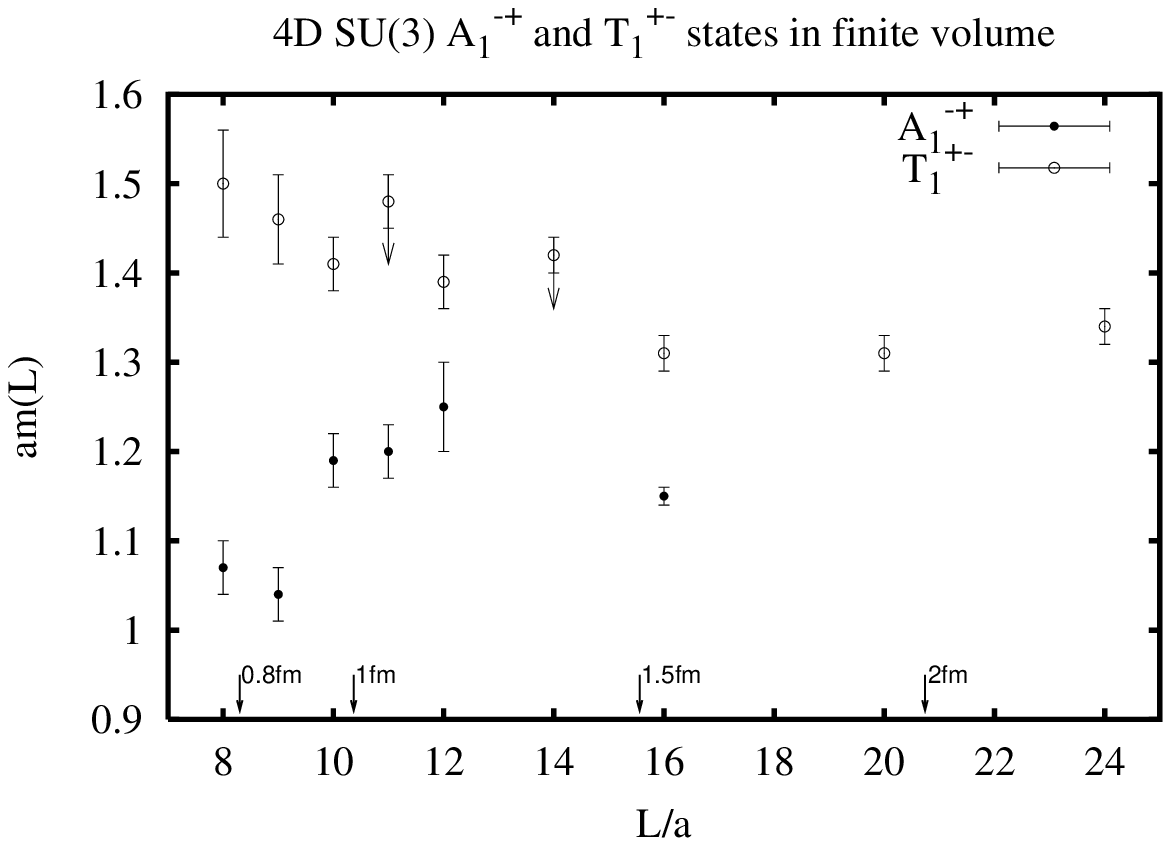,angle=0,width=15cm}
        \psfig{file=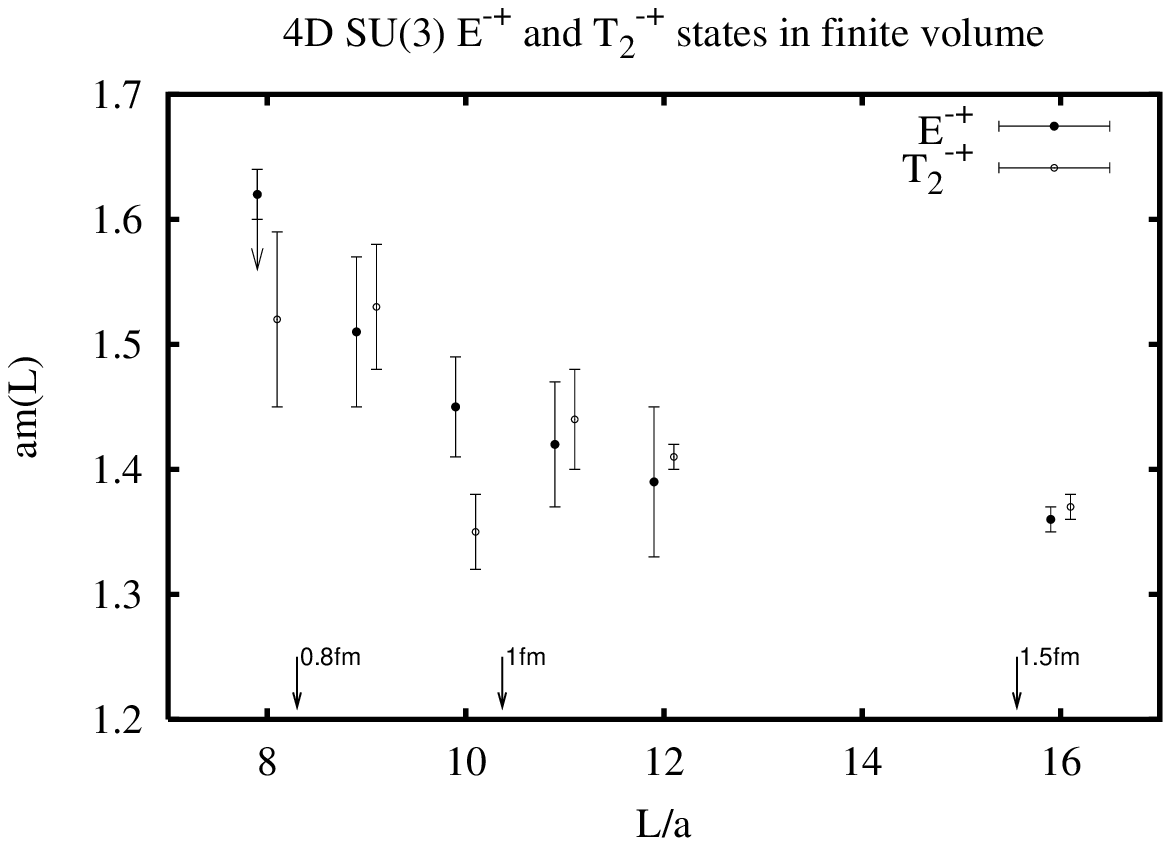,angle=0,width=15cm}
            \end{minipage}}
\caption[a]{Volume dependence of some light  states in various representations.}
\la{fig:other_g}
\end{figure}
\clearpage
\clearpage
\begin{figure}[t]
\centerline{\begin{minipage}[c]{15cm} 
\psfig{file=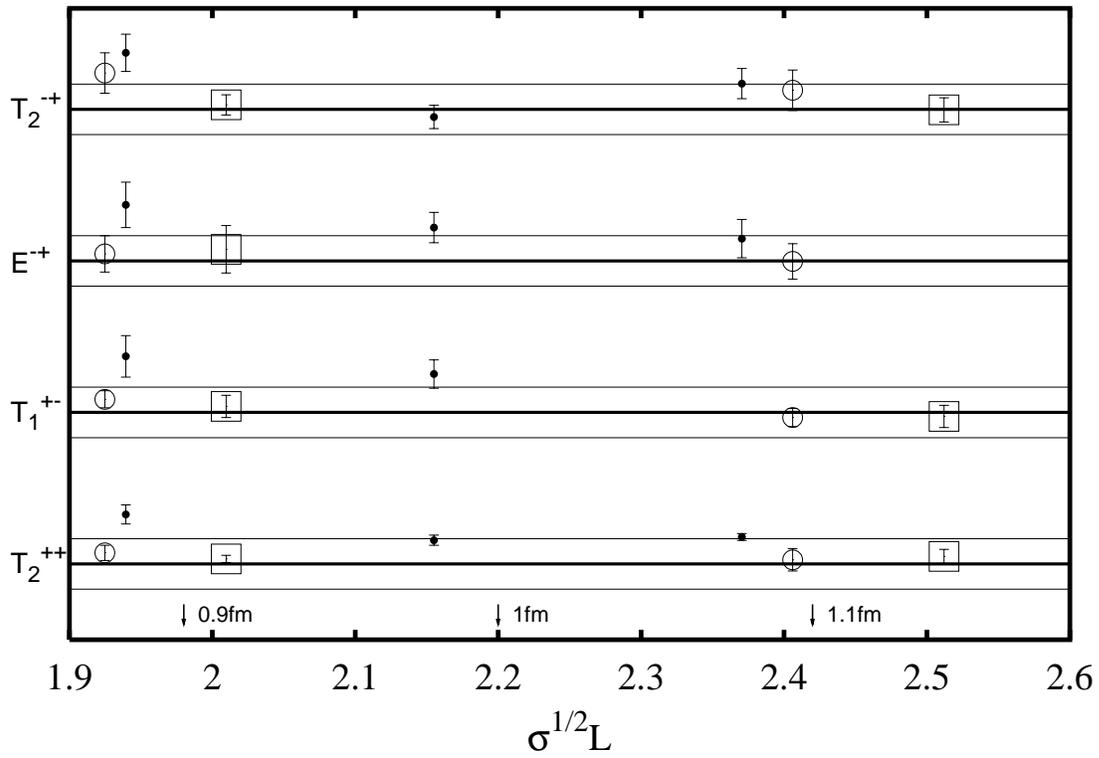,angle=0,width=15cm} 
\psfig{file=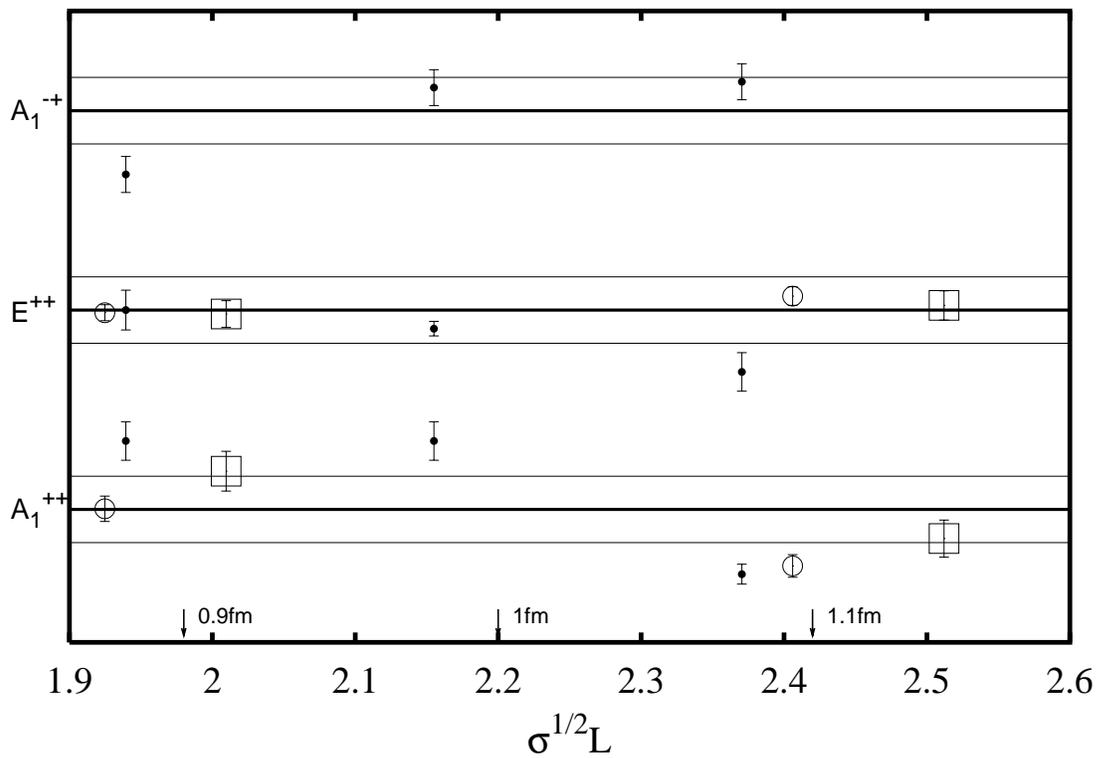,angle=0,width=15cm} 
\end{minipage}}
\caption[a]{Finite-volume mass shifts $m(L)/m(L=\infty)-1$ 
in various representations; 
SU(3)$\rightarrow$~black~circles,
SU(4)$\rightarrow$~white~circles,
SU(6)$\rightarrow$~white~squares. The thin lines indicate the $\pm5\%$ levels of 
variation.}
\la{fig:other_rep}
\end{figure}
\end{document}